\documentclass[prd,twocolumn]{revtex4}

\usepackage{bm}
\usepackage{amssymb}
\usepackage{graphicx}
\usepackage{epstopdf}
\usepackage{amsfonts}
\usepackage[figuresright]{rotating}  
\usepackage{amssymb}
\usepackage{amsmath}
\usepackage{url}

\begin{document}

\title{The CBE Hardware Accelerator for Numerical Relativity: A Simple Approach}

\author{Gaurav Khanna} 

\address{Physics Department, University of Massachusetts at Dartmouth\\
285 Old Westport Rd, Dartmouth, Massachusetts 02747, USA\\
gkhanna@umassd.edu}

\begin{abstract}
Hardware accelerators (such as the Cell Broadband Engine) have recently received a 
significant amount of attention from the computational science community 
because they can provide significant gains in the overall performance 
of many numerical simulations at a low cost. However, such accelerators 
usually employ a rather unfamiliar and specialized programming model that often requires 
advanced knowledge of their hardware design. In this article, we demonstrate an alternate 
and simpler approach towards managing the main complexities in the programming 
of the Cell processor, called {\it software caching}. We apply this technique 
to a numerical relativity application: a time-domain, finite-difference 
Kerr black hole perturbation evolver, and present the performance results. 
We obtain gains in the overall performance of generic simulations that are close to 
the theoretical maximum that can be obtained through our parallelization approach.      
\end{abstract}


\maketitle

\section{Introduction}	
Computer simulations are playing an increasingly important role in nearly every area of 
science and engineering today. The main drive behind this trend is the rapid increase in 
the overall performance of computer hardware over the past several decades ({\it Moore's Law})
and its relatively low cost. 

It is interesting to note that the overall performance of certain specific computing technologies, such as 
graphics cards, gaming consoles, etc. has continued to increase at a rate much higher than that 
of traditional workstation processors, thus making scientific computing on such devices an intriguing  
possibility: Compute Unified Device Architecture (CUDA)~\cite{cuda} is NVIDIA's general-purpose software 
development system for graphics processing units (GPUs); Cell Broadband Engine (CBE)~\cite{cbe} is a 
processor that was designed by a collaboration between Sony, Toshiba, and IBM and is being used in gaming 
consoles (Sony's Playstation~\cite{ps3}) as well as high-performance computing hardware (IBM's Cell 
blades~\cite{blades}, LANL RoadRunner~\cite{roadrunner}). 

In this article, we will focus entirely on the CBE and demonstrate its use in the acceleration 
of scientific computing applications. In particular, we take a sample application from the 
numerical relativity (NR) community -- a Teukolsky equation solver~\cite{teuk} which is a 
finite-difference (2+1)D linear, hyperbolic, homogeneous partial differential equation (PDE) 
solver -- and implement the low-level parallelism that is offered by the CBE architecture. 
We describe the approach taken and its outcome in detail. This NR application is fairly generic, 
i.e. it is of a type that is quite common in various fields of science and engineering. Therefore 
our work would be of interest to a larger community of computational scientists. 

It is worth pointing out that various researchers have performed a similar evaluation of the CBE using 
other finite-difference PDE solvers~\cite{other} with very promising results. These studies typically 
involved extensive coding at a level that requires rather specialized and advanced knowledge of the CBE's 
design. Indeed, such an investment is very necessary if the goal is to extract the maximum amount of performance 
that the processor can offer. However, our approach in this article is one in which we will make a considerably 
smaller investment in the specialized coding and yet obtain strong gains in overall application performance.
Our approach makes use of a {\it software caching} mechanism on the Cell's Synergistic Processing Elements (SPEs) 
that was made available recently in IBM's Cell SDK~\cite{sdk}. We describe this less-known mechanism and our 
implementation in detail, later in this article. 

\section{Numerical Relativity}
Several gravitational wave observatories~\cite{ligo} are currently being built all over the world: LIGO in the United 
States, GEO/Virgo in Europe and TAMA in Japan. These observatories will open a new window onto the Universe by 
enabling scientists to make astronomical observations using a completely new medium -- gravitational waves (GW), 
as opposed to electromagnetic waves (light). These waves were predicted by Einstein's relativity theory, but have 
not been directly observed because the required experimental sensitivity was simply not advanced enough, until 
very recently.

Numerical relativity is an area of computational science that emphasizes the detailed modeling of strong sources 
of GWs -- collisions of compact astrophysical objects, such as neutron stars and black holes. Thus, it plays an 
extremely important role in the area of GW astronomy and gravitational physics, in general. Moreover, the NR 
community has also contributed to the broader computational science community by developing an open-source, modular, 
parallel computing infrastructure called {\it Cactus}~\cite{cactus}.

The specific NR application we have chosen for consideration in this work is one that evolves the perturbations 
of a rotating (Kerr) black hole, i.e. solves the Teukolsky equation~\cite{teuk} in the time-domain. This equation is 
 essentially a linear wave-equation in Kerr space-time geometry. The next two subsections provide more detailed 
information on this equation and the associated numerical solver code.

\subsection{Teukolsky Equation}
The Teukolsky master equation describes scalar, vector and tensor field perturbations in the space-time of
Kerr black holes~\cite{eqn}. In Boyer-Lindquist coordinates, this equation takes the form
\begin{eqnarray}
\label{teuk0}
&&
-\left[\frac{(r^2 + a^2)^2 }{\Delta}-a^2\sin^2\theta\right]
         \partial_{tt}\Psi
-\frac{4 M a r}{\Delta}
         \partial_{t\phi}\Psi \nonumber \\
&&- 2s\left[r-\frac{M(r^2-a^2)}{\Delta}+ia\cos\theta\right]
         \partial_t\Psi\nonumber\\  
&&
+\,\Delta^{-s}\partial_r\left(\Delta^{s+1}\partial_r\Psi\right)
+\frac{1}{\sin\theta}\partial_\theta
\left(\sin\theta\partial_\theta\Psi\right)+\nonumber\\
&& \left[\frac{1}{\sin^2\theta}-\frac{a^2}{\Delta}\right] 
\partial_{\phi\phi}\Psi +\, 2s \left[\frac{a (r-M)}{\Delta} 
+ \frac{i \cos\theta}{\sin^2\theta}\right] \partial_\phi\Psi  \nonumber\\
&&- \left(s^2 \cot^2\theta - s \right) \Psi = 0  ,
\end{eqnarray}
where $M$ is the mass of the black hole, $a$ its angular momentum per unit mass, $\Delta = r^2 - 2 M r + a^2$ and 
$s$ is the ``spin weight'' of the field. The $s = \pm 2$ versions of these equations describe the radiative degrees 
of freedom of the gravitational field, and thus are the equations of interest here. As mentioned previously, this equation 
is an example of linear, hyperbolic, homogeneous PDEs which are quite common in several areas of science and 
engineering, and can be solved numerically using a variety of finite-difference schemes.  

\subsection{Teukolsky Code}
Ref.~\cite{klpa} demonstrated stable numerical evolution of Eq.\ (\ref{teuk0}) for $s=-2$ using the well-known 
Lax-Wendroff numerical evolution scheme. Our Teukolsky code uses the exact same approach, therefore the contents of 
this section are largely a review of the work presented in the relevant literature~\cite{klpa}. 

Our code uses the tortoise coordinate $r^*$ in the radial direction and azimuthal coordinate $\tilde{\phi}$. 
These coordinates are related to the usual Boyer-Lindquist coordinates by
\begin{eqnarray}
dr^* &=& \frac{r^2+a^2}{\Delta}dr 
\end{eqnarray}
and
\begin{eqnarray}
d\tilde{\phi} &=& d\phi + \frac{a}{\Delta}dr \; . 
\end{eqnarray}  
Following Ref.~\cite{klpa}, we factor out the azimuthal dependence and use the ansatz,
\begin{eqnarray}
\label{eq:psiphi}
\Psi(t,r^*,\theta,\tilde{\phi}) &=& e^{im\tilde{\phi}} r^3 \Phi(t,r^*,\theta) .
\end{eqnarray}
Defining
\begin{eqnarray}
\Pi &\equiv& \partial_t{\Phi} + b \, \partial_{r^*}\Phi \; , \\
b & \equiv &
\frac { {r}^{2}+{a}^{2}}
      { \Sigma} \; , 
\end{eqnarray}
and
\begin{eqnarray}
\Sigma^2 &\equiv &  (r^2+a^2)^2-a^2\,\Delta\,\sin^2\theta
\; 
\label{pi_eq}
\end{eqnarray} 
allows the Teukolsky equation to be rewritten as
\begin{eqnarray}
\label{eq:evln}
\partial_t \mbox{\boldmath{$u$}} + \mbox{\boldmath{$M$}} \partial_{r*}\mbox{\boldmath{$u$}} 
+ \mbox{\boldmath{$Lu$}} + \mbox{\boldmath{$Au$}} =  0 ,
\end{eqnarray}
where 
\begin{equation}
\mbox{\boldmath{$u$}}\equiv\{\Phi_R,\Phi_I,\Pi_R,\Pi_I\}
\end{equation}
is the solution vector. The subscripts $R$ and $I$ refer to the real
and imaginary parts respectively (note that the Teukolsky function
$\Psi$ is a complex valued quantity). Explicit forms for the matrices {\boldmath{$M$}},
{\boldmath{$A$}} and {\boldmath{$L$}} can be easily found in the relevant literature~\cite{klpa}.  
Rewriting Eq.\ (\ref{eq:evln}) as 
\begin{equation}
\partial_t \mbox{\boldmath{$u$}} + \mbox{\boldmath{$D$}}
\partial_{r^*} \mbox{\boldmath{$u$}}
=  \mbox{\boldmath{$S$}}\; , 
\label{new_teu2}
\end{equation}
where
\begin{equation}
 \mbox{\boldmath{$D$}} \equiv \left(\begin{matrix}
                    b &   0   &  0  &  0 \cr
                    0  &   b   &  0  &  0 \cr
                    0  &   0   &  -b  &  0 \cr
                    0  &   0   &  0  &  -b \cr
                \end{matrix}\right),
\label{d_matrix}
\end{equation}
\begin{equation}
\mbox{\boldmath{$S$}} = -(\mbox{\boldmath{$M$}} - \mbox{\boldmath{$D$}})
\partial_{r^*}\mbox{\boldmath{$u$}}
- \mbox{\boldmath{$L$}}\mbox{\boldmath{$u$}} 
- \mbox{\boldmath{$A$}}\mbox{\boldmath{$u$}},
\end{equation}
and using the Lax-Wendroff iterative scheme, we obtain stable evolutions.
Each iteration consists of two steps: In the first step, the solution vector 
between grid points is obtained from
\begin{eqnarray}
\label{lw1}
\mbox{\boldmath{$u$}}^{n+1/2}_{i+1/2} &=& 
\frac{1}{2} \left( \mbox{\boldmath{$u$}}^{n}_{i+1}
                  +\mbox{\boldmath{$u$}}^{n}_{i}\right)
- \\
&  &\frac{\delta t}{2}\,\left[\frac{1}{\delta r^*} \mbox{\boldmath{$D$}}^{n}_{i+1/2}
  \left(\mbox{\boldmath{$u$}}^{n}_{i+1}
                  -\mbox{\boldmath{$u$}}^{n}_{i}\right)
- \mbox{\boldmath{$S$}}^{n}_{i+1/2} \right] \; .\nonumber
\end{eqnarray}
This is used to compute the solution vector at the next time step,
\begin{equation}
\mbox{\boldmath{$u$}}^{n+1}_{i} = 
\mbox{\boldmath{$u$}}^{n}_{i}
- \delta t\, \left[\frac{1}{\delta r^*} \mbox{\boldmath{$D$}}^{n+1/2}_{i}
  \left(\mbox{\boldmath{$u$}}^{n+1/2}_{i+1/2}
                  -\mbox{\boldmath{$u$}}^{n+1/2}_{i-1/2}\right)
- \mbox{\boldmath{$S$}}^{n+1/2}_{i} \right] \, .
\label{lw2}
\end{equation}
The angular subscripts are dropped in the above equation for clarity. All angular
derivatives are computed using second-order, centered finite difference expressions. 

Following Ref.~\cite{klpa}, we set $\Phi$ and $\Pi$ to zero on the inner
and outer radial boundaries. Symmetries of the spheroidal harmonics
are used to determine the angular boundary conditions: For even $|m|$
modes, we have $\partial_\theta\Phi =0$ at $\theta = 0,\pi$ while $\Phi =0$ at 
$\theta = 0,\pi$ for modes of odd $|m|$.
 
\begin{figure}[th]
%\centerline{\psfig{file=qnm.eps,width=10cm}}
\centerline{\includegraphics[width=8cm]{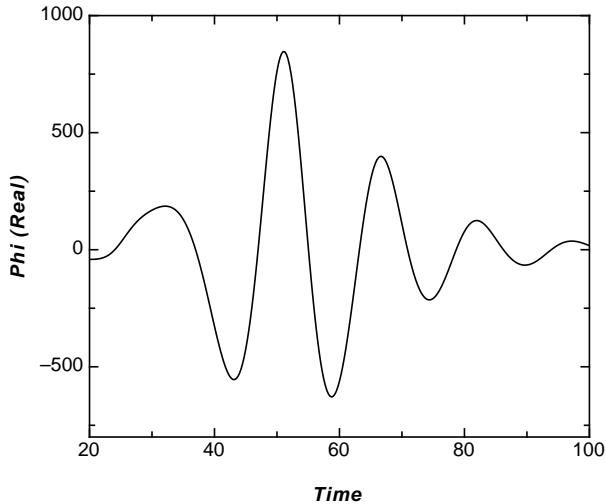}}
\vspace*{8pt}
\caption{The quasi-normal ringing for a black hole with
$a/M = 0.9$; the $l = 0$, $m = 0$ mode is shown here. The evolution 
of the real part of the Teukolsky function, extracted at $r = 20 M$, 
$\theta = \pi/2$. \label{qnm}}
\end{figure}

As a sample numerical result, we take initial data corresponding to a narrow 
Gaussian pulse.  This data perturbs the black hole, causing it to ring down according 
to its characteristic quasi-normal frequencies. Fig.~\ref{qnm} shows the results, 
illustrating the quasi-normal ringing for the $l = 0$, $m = 0$ 
mode of a black hole with spin parameter $a = 0.9$.

\section{Cell Broadband Engine}
The CBE is a completely new processor, that was developed collaboratively by
Sony, IBM and Toshiba primarily for multimedia applications. This processor has a
general purpose (PowerPC) CPU, called the PPE (that can execute two (2) software threads
simultaneously) and eight (8) special-purpose compute engines, called SPEs only   
for numerical computation. Each SPE can perform vector operations, which implies
that it can compute on multiple data, in a single instruction (SIMD). All of these
compute elements are connected to each other through a high-speed interconnect bus 
(EIB). A single 3.2 GHz CBE (original - 2006/2007) has a peak performance of over 
200 GFLOP/s in single-precision floating point computation and 15 GFLOP/s in 
double-precision. The current (2008) release of the CBE, called the {\it PowerXCell}, 
has design improvements that bring the double-precision performance up to 100 GFLOP/s. 

The main programming challenge introduced by this new design, is that one has 
to explicitly manage the memory transfer between the PPE and the SPEs. The
PPE and SPEs are equipped with a DMA engine -- a mechanism that enables data
transfer to and from main memory and each other. Now, the PPE can access main
memory directly, but the SPEs can only directly access their own, rather limited
(256KB) local-store. This poses a challenge for many applications,
including the NR application that we are considering in this article. However, 
the compilers in the most recent release (3.1) of IBM's Cell SDK enable a 
{\em software caching} mechanism that allows for the use of the SPE local-store 
as a conventional cache, thus negating the need of transferring data manually 
from main memory. In essence, this mechanism allows one to write SPE programs that 
can access variables stored in the PPE's address space (IBM refers this to as 
{\it effective address space} support). From our viewpoint, this feature takes 
away the main challenge associated to programming the CBE, and thus considerably 
reduces the amount of specialized code development needed. Therefore, we will take 
this approach in the development of a CBE optimized version 
of our NR application. Of course, it is worth pointing out that this approach 
will not allow the processor to perform at full potential; thus one should not 
expect to achieve the maximal performance gain that would be possible otherwise.

Another important mechanism that allows communication between the the different 
elements (PPE, SPEs) of the CBE is the use of mailboxes. These are special purpose 
registers that can be used for uni-directional communication. Each SPE has three (3) 
mailboxes -- two (2) outbound, that can hold only a single entry, and one (1) inbound, that 
can hold four (4) entries. These are typically used for synchronizing the computation 
across the SPEs and the PPE, and that is primarily how we made use of these registers 
as well. Details on our specific use of these various aspects of the CBE for the 
NR application appear in the next section of this article.

\section{CBE Teukolsky Code Implementation}
As mentioned in the previous section, our approach towards running code on the 
Cell's SPEs involves making use of the {\it effective address space} support. In 
this manner we avoid the involved programming associated to explicitly managing 
the data exchange between the PPE and the SPEs. Therefore, once we have decided 
upon the portion of the Teukolsky code that would be appropriate to run on the 
SPEs, the rest would be straightforward. 

Since the SPEs are the main compute engines of the CBE, one would want them to execute 
the most compute intensive tasks of a code in a data- or task- parallel fashion, 
and leave the rest (input/output, synchronization, etc.) to run on the PPE. We employ 
a data-parallel model, which is straightforward to implement in a code like ours -- 
we simply perform a domain decomposition of our finite-difference numerical grid 
and allocate the different parts of the grid to different SPEs. In particular, we 
parallelize along the $r^*$ coordinate dimension, because that typically has two 
orders-of-magnitude more grid points than the other ($\theta$) dimension. 

Upon performing a basic profiling of our code using the GNU profiler {\bf gprof}, 
we learn that the computing the ``right-hand-sides'' of the Lax-Wendroff steps 
i.e. the quantities within the square-brackets of Eqs.\ (\ref{lw1}) and 
(\ref{lw2}), take nearly {\it 75\%} of the application's overall runtime. Thus, it 
is natural to consider accelerating this ``right-hand-side'' computation using 
data-parallelization on the SPEs. We anticipate that this observation is fairly 
typical for codes of this type.

Now, to move this computation to the SPEs, we simply take  
the corresponding PPE code and copy it into a skeleton SPE code and then add in 
the appropriate declarations for all the field arrays such that they point to the 
corresponding quantities in the PPE's address space. To be more explicit, lets say 
that we have a field array quantity {\bf phi} defined in the PPE portion of the code 
as follows,
\begin{verbatim}
double phi[M][N];
\end{verbatim}
where $M$ and $N$ are the dimensions of the array in the $\theta$ and $r^*$ directions 
respectively. To point to this exact same array from the SPE code, we simply need to 
declare it as,
\begin{verbatim}
extern __ea double phi[M][N];
\end{verbatim}
where the {\bf \_\_ea} type qualifier is a PPE address namespace identifier. That 
entirely takes care of the complicated matter of transferring the necessary data 
back-and-forth between the SPEs and the PPE. Each SPE maintains a cache (size can be 
controlled at compile time using the {\bf -mcache-size} flag) in its local-store, wherein 
it stores a small fraction of the whole array for immediate use; therefore its 
functioning is identical to the traditional hardware caches found in all modern 
processors. 

Finally, as far as making use of the SIMD capabilities of the SPEs are concerned, we simply 
enable the auto-vectorizing capabilities of the {\bf spu-gcc} compiler, as opposed to 
performing those optimizations by hand. The only other issue remaining, is that of 
synchronizing the PPE and the SPEs, which is easily done by using mailboxes as mentioned 
earlier. 

\section{Performance Results}
Recall that the ``right-hand-side'' computation, that we are coding for parallelized  
execution on the Cell's SPEs takes nearly 75\% of the overall runtime of the application. 
This means that at best, we can expect an acceleration of a factor of four (4) in the 
application's overall runtime ({\it Amdahl's law}). In this section, we compare our CBE 
accelerated code's performance relative to this maximum achievable gain. 

\begin{figure}[th]
%\centerline{\psfig{file=perf.eps,width=10cm}}
\centerline{\includegraphics[width=9cm]{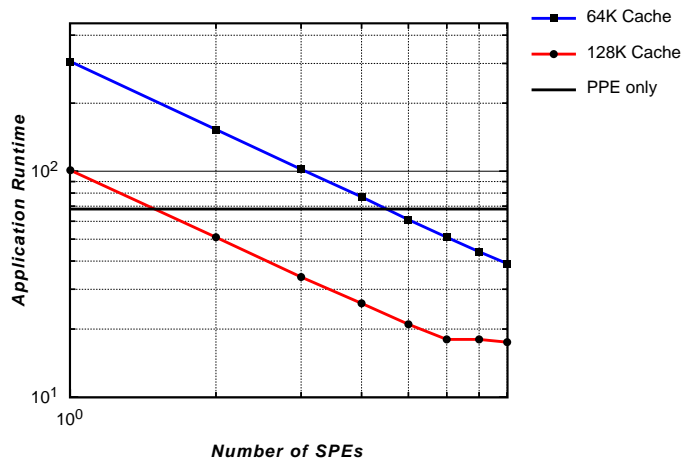}}
\vspace*{8pt}
\caption{The overall application runtime of our CBE Teukolsky code, as a function of the
number of SPEs used, for two (2) different software cache sizes. The performance gains owing 
to parallelization are significant, with more gain for larger cache size. We also depict the 
runtime of the PPE-only code for comparison purposes. \label{perf}}
\end{figure}

Fig.~\ref{perf} depicts the total application runtime as a function of the number of 
SPEs used, for two (2) different values of the software cache. One can clearly see that   
although the performance gain using this approach is highly dependent on the size 
of the software cache used, the gains are substantial. For the 
case of the $128K$ cache size, the maximum acceleration of the application is a factor of 
$3.885$ -- this translates to {\it over 97\% of the maximum theoretical gain} possible 
through our parallel approach!

Another observation that can be made from the $128K$ cache data plotted in Fig.~\ref{perf} 
is that after six (6) SPEs are in use, not much is gained by including additional SPEs. 
The reason for this is simply that even with only six (6) SPEs in use, the runtime associated 
to the ``right-hand-side'' computation becomes negligible as a fraction of the application's 
overall runtime. Thus, including additional SPEs to further accelerate that computation 
yields no benefit. If we wanted to accelerate the application even further, we would need 
to further port some of the other PPE-based computations onto the SPEs.

It is worth pointing out that these tests were performed on an IBM QS21 blade server 
that is equipped with the original (2006/2007) CBE -- the one with the rather limited 
double-precision floating-point performance. All of our codes were run using double-precision 
floating-point accuracy because that is the common practice in the NR community and 
also a necessity for finite-difference based evolutions, especially if a large number 
of time-steps are involved. On the current {\it PowerXCell} processor, the overall performance 
of our code is likely to be much better.  

\section{Conclusions}
We demonstrate that it is possible to obtain a significant performance gain 
(upto 97\% of theoretical maximum) on a typical finite-difference PDE solver by making use 
of the SPEs of the Cell processor, via a programming model that is relatively very 
straightforward to implement. The approach involves the use of {\it software caching} 
through the recently made available {\it effective address space} support on the Cell's 
SPEs. 

By making use of this approach many scientific applications may obtain high accelerations
on CBE hardware with minimal investment in detailed and specialized coding. Moreover, it is 
worth noting that we obtain these very strong results from the the original release of the Cell 
processor -- the one found in the Sony PS3 -- therefore, one can obtain these performance 
gains using such extremely low-cost hardware. 

\section*{Acknowledgments}
We thank Jens Brietbart for several helpful suggestions relating to this work. GK would like 
to acknowledge support from the National Science Foundation (NSF grant numbers: PHY-0831631, PHY-0902026).

\end{document}